\documentclass[12pt]{iopart}
\usepackage{iopams}  
\usepackage{graphicx}
\usepackage{dcolumn}
\usepackage{bm}
\usepackage{color}

\newcommand{\exciting}{{\usefont{T1}{lmtt}{b}{n}exciting}}
\begin{document}

\title{Robust mixing in self-consistent linearized augmented planewave calculations}
\author{Jongmin Kim, Andris Gulans, Claudia Draxl}
\address{Institut f{\"u}r Physik and IRIS Adlershof, Humboldt-Universit{\"a}t zu Berlin, 12489 Berlin, Germany}
\ead{jongmin.kim@physik.hu-berlin.de, andris.gulans@physik.hu-berlin.de, claudia.draxl@physik.hu-berlin.de}

\begin{abstract}
We devise a mixing algorithm for full-potential (FP) all-electron calculations in the linearized augmented planewave (LAPW) method. Pulay's direct inversion in the iterative subspace is complemented with the Kerker preconditioner and further improvements to achieve smooth convergence, avoiding charge sloshing and noise in the exchange-correlation potential. As the Kerker preconditioner was originally designed for the planewave basis, we have adapted it to the FP-LAPW method and implemented in the \exciting\ code. Applications to the $2\times 2$ Au(111) surface with a vacancy and to the Pd(111) surface demonstrate that this approach and our implementation work reliably with both density and potential mixing. 
\end{abstract}

\vspace{2pc}

\maketitle

\section{Introduction}
Density-functional theory (DFT) \cite{Hohenberg1964, Kohn1965} is used for a wide range of problems that cover electronic, mechanical, and vibrational properties of atoms, molecules, and solids. The diversity of applications to various materials requires that numerical algorithms employed in electronic-structure codes are robust. Otherwise, a poorly chosen or implemented algorithm may lead to unnecessary long computation times as well as to  non-converging calculations \cite{Woods2019}. Such issues hinder not only individual calculations but also are particularly {harmful for high-throughput} studies. 

One particular problem commonly recurring in electronic-structure theory is convergence of the self-consistent Kohn-Sham equation:
\begin{equation}
\label{eq:KS}
\left[-\frac{1}{2}\nabla^2+v_{\mathrm {KS}}(\bf r)\right] \psi_{\mathrm i\bf{k}}(\bf r)=\epsilon_{\mathrm i\bf{k}} \psi_{\mathrm i\bf{k}}(\bf r),
\end{equation}
where $\epsilon_{\mathrm i\bf{k}}$, $v_{\mathrm {KS}}(\bf r)$, and $\psi_{\mathrm i\bf{k}}(\bf r)$ represent Kohn-Sham eigenenergies, potential, and wavefunctions, respectively. $v_{\mathrm {KS}}(\bf r)$ depends on the electron density:
\begin{equation}
\label{eq:density}
\rho (\bf r)=\sum_{\mathrm \bf{k}} \omega_{\mathrm \bf{k}} \sum_{\mathrm i} \it f_{\mathrm i\bf{k}} \left|\psi_{\mathrm i\bf{k}}(\bf r)\right|^{\mathrm 2},
\end{equation}
where $\omega_{\bf{k}}$ is the weight of the $\bf k$-point and $f_{\mathrm i\bf{k}}$ the occupation factor. Eq.~\ref{eq:KS} defines a non-linear eigenvalue problem. To solve it, one considers a linear problem with a fixed potential $v^\mathrm{in}_{\mathrm {KS}}(\bf r)$. After computing $\psi_{\mathrm i\bf{k}}(\bf r)$ and subsequently calculating $\rho (\bf r)$, one obtains a new potential $v^\mathrm{out}_{\mathrm {KS}}(\bf r)$ that serves, in principle, as $v^\mathrm{in}_{\mathrm {KS}}(\bf r)$ for the next iteration of the KS equation (Eq.~\ref{eq:KS}). This procedure is repeated until self-consistency is reached, namely, $v^\mathrm{in}_{\mathrm {KS}}({\bf r}) \approx  v^\mathrm{out}_{\mathrm {KS}}(\bf r)$.

In practice, at every step, the potential is constructed as a mixture (linear combination) of $v^\mathrm{out}_{\mathrm {KS}}(\bf r)$ and {\it history} of $v^\mathrm{in}_{\mathrm {KS}}(\bf r)$. One can likewise target self-consistency in the electron density instead of the potential, applying a corresponding mixing scheme. Regardless of the choice, the success and efficiency of this procedure depends on how exactly either of these two quantities is updated, and the problem resembles a multivariate optimization. As a result, a number of mixing methods~{\cite{Broyden1965,Johnson1988,Pulay1980,Pulay1982,Gonze1996, Bowler2000, Fang2009, Pratapa2015, Banerjee2016}} relies on standard optimization techniques or is intimately related to them. Despite successful applications to numerous systems, there are challenging cases that require further improvement of these methods. In particular, metallic systems with large unit cells suffer from an instability known as charge sloshing~\cite{Woods2019, Anglade2008, Kerker1981, Kresse19961, Kresse19962, Kohyama1996, Shiihara2008, Sundararaman2017}. This issue originates from (i) a constant non-zero susceptibility, $\chi$, of metallic systems at small wave-vectors and / or (ii) a factor of $G^{-2}$ in the Hartree potential at small \textbf{G} components. It can, e.g., be seen in the error of the output potential 
\begin{equation}
\label{eq:errorpotential}
\delta v^\mathrm{out}_{\mathrm {KS}}({\bf {G}})= \sum\limits_{{\bf G}} \frac{4\pi}{{\bf {G}}^{2}} \ \chi(\left|\bf {G}\right|) \ \delta v^\mathrm{in}_{\mathrm {KS}}({\bf {G}}).
\end{equation}
that indicates that a small error in the input potential for short \textbf{G} is amplified due to these two factors.

Several studies have offered a solution to this problem~\cite{Anglade2008, Kerker1981, Kresse19961, Kresse19962, Raczkowski2001, Vanderbilt1984, Ho1982}. Among them, the Kerker preconditioner is widely used to reduce the charge-sloshing instability induced at small \textbf{G} (long-wavelength) \cite{Kerker1981}, thus making the self-consistency procedure stable for metallic systems with large unit cells.

Mixing schemes are most frequently developed with planewave / pseudopotential methods in mind. Nevertheless, one can adopt the same techniques for all-electron implementations. For instance, {variants of multisecant Broyden method have} been successfully applied in FP calculations with the LAPW basis set {\cite{Marks2008, Marks2013}. Corresponding implementations have} become the default choice in two FP-LAPW codes, i.e. \exciting~\cite{Gulans2014} and Wien2k~\cite{Blaha1990}. Our experience shows that this method, especially a multisecant Broyden type-1 method, which estimates a Jacobian \cite{Marks2008}, performs well in most cases, but is unstable with respect to charge-sloshing. Unfortunately, also the aforementioned solutions to this problem target planewave implementations. Mixing potentials rather than densities, pose additional numerical difficulties. For instance, the generalized-gradient approximation (GGA) for the exchange-correlation functional may lead to noisy potentials in the low-density regions, making the problem of finding a self-consistent potential ill-conditioned. 

In this paper, we tackle the issues of charge-sloshing and noisy potentials in FP-LAPW calculations. We implement the Kerker scheme and a modified Pulay mixer in the \exciting\ code and discuss their performance with two benchmark systems: (i) the $2\times 2$ Au(111) surface with a vacancy and (ii) the Pd(111) surface. We show that the modified Pulay mixer with the Kerker preconditioner is robust for both density and potential mixing. 

\section{Kerker mixing \label{sec:kerker}}

Considering the $i$-th step of the self-consistent field procedure, the $(i+1)$-st guess of the Kohn-Sham potential can be obtained via a simple linear relation:
\begin{equation}
\label{eq:linear}
v^\mathrm{in}_{i+1}(\mathbf{r}) = v^\mathrm{in}_i(\mathbf{r}) + \alpha \left[ v^\mathrm{out}_{i}({\bf r}) - v^\mathrm{in}_{i}({\bf r}) \right],
\end{equation}
where $\alpha$ is some prefactor. This simple approach may work in many cases, however, it fails in calculations of metallic systems with a large unit cell. The linear-response properties in these systems are such that a small change in $v^\mathrm{in}_i$ triggers a large change in $v^\mathrm{out}_i$ at short wavevectors $\bf G$  as outlined above. Such an instability causes massive charge fluctuations over the self-consistency cycle which is known as charge sloshing. These short-$\mathbf{G}$ fluctuations can be suppressed using the Kerker scheme~\cite{Kerker1981}. In a planewave basis, the mixing relation in Eq.~\ref{eq:linear} transforms into
\begin{equation}
\label{eq:kerker}
v^\mathrm{in}_{i+1}({\bf G}) = v^\mathrm{in}_i ({\bf G}) + \alpha\frac{G^2}{G^2+\lambda^2} \left[ v^\mathrm{out}_{i}({\bf G}) - v^\mathrm{in}_{i}({\bf G}) \right],
\end{equation}
where $\lambda$ is a parameter that determines the range of ${\bf G}$ over which changes of the potential are suppressed.

Eq.~\ref{eq:kerker} is straightforward to implement in the planewave methods, but not as simple in others. To describe its implementation in the FP-LAPW method, we briefly sketch its characteristics. In this method, the unit cell is divided into the interstitial region (\textit{I}) and atomic spheres (commonly known as muffin-tin spheres), that are centered at the atomic positions. The potential (as is the density) is expanded in terms of planewaves in $I$ and in spherical harmonics in the atomic spheres (labelled $\alpha$), respectivley:
\begin{equation}
\label{eq:LAPWpotential}
v({\bf r}) = 
\cases{\sum\limits_{ {\bf G}} v_{I}({\bf G}) \ e^{i{\bf G}{\bf r}} & ${\bf r}\, {\in I}$\\ \sum\limits_{lm} v^{\alpha}_{lm}({\bf r}) \ Y_{lm}({\bf \widehat{r}}) & ${\bf r}\, {\in \mathrm{MT}_{\alpha}}$.\\}
\end{equation}
Since performing a Fourier transform is not feasible in this case, calculating $v^\mathrm{in}_{i+1}$ directly from Eq.~\ref{eq:kerker} is not practical.

To make Eq.~\ref{eq:kerker} applicable in the FP-LAPW case, we transform it to the real space and obtain
\begin{equation}
\label{eq:kerkerRS}
v^\mathrm{in}_{i+1}(\mathbf{r}) = v^\mathrm{in}_i(\mathbf{r}) +\alpha \left[ 1+\lambda^2\left(\nabla^2-\lambda^2 \right)^{-1} \right] \left[v^\mathrm{out}_{i}({\bf r}) - v^\mathrm{in}_{i}({\bf r})\right].
\end{equation}
The operator $\hat{K}=\left(\nabla^2-\lambda^2 \right)^{-1}$ cannot be applied to a function directly. {Instead, we calculate $V(\mathbf{r})=\hat{K}f(\mathbf{r})$ as a solution of the screened Poisson equation:}
%
\begin{equation}
\label{eq:screened}
\left(\nabla^2-\lambda^2 \right)V(\mathbf{r})=f(\mathbf{r}),
\end{equation}
{where $f(\mathbf{r})$ is some function. If we set $f(\mathbf{r})=-4\pi\rho(\mathbf{r})$, $V(\mathbf{r})$ corresponds to the potential due to the charge density $\rho(\mathbf{r})$ assuming the screened Coulomb interaction. Keeping in mind that our goal is merely to apply the operator $\left(\nabla^2-\lambda^2 \right)^{-1}$ on a given function, we treat Eq.~\ref{eq:screened} as if we had an electrostatics problem in which the input $\rho(\mathbf{r})=-\left[v^\mathrm{out}_{i}({\bf r}) - v^\mathrm{in}_{i}({\bf r})\right]/4 \pi$ is given. 
From this, we can then obtain $V(\mathbf{r})$.}

Eq.~\ref{eq:screened} can be solved in the spirit of the pseudo-charge method suggested by Weinert \cite{Weinert1981}. It was originally proposed for calculating the Hartree potential in FP-LAPW calculations, but the same idea can be exploited for solving the screened Poisson equation as implemented in Ref.~\cite{Tran2011}. The original purpose of the algorithm presented in that study was an implementation of the screened exchange in a hybrid exchange-correlation functional but it also meets our needs. The crucial steps of the method are described below.

At first, we calculate the screened potential in the interstitial region. To do so, the charge density $\rho(\mathbf{r})$ is replaced by a smooth pseudo-charge density $\bar{\rho}(\mathbf{r})$, for which it is easy to perform a Fourier transform. Note that the input \textit{charge density} is given in the same form as the potential in Eq.~\ref{eq:LAPWpotential}. The pseudo-charge density is constructed as
\begin{equation}
\label{eq:pseudodensity}
\bar{\rho}(\mathbf{r})=\sum\limits_{\mathbf{G}}\rho_\mathbf{G} \ e^{i\mathbf{G}\mathbf{r}}+\sum\limits_{\alpha} \tilde{\rho}^\alpha(\mathbf{r}),
\end{equation}
where the first term is the given interstitial density. Note that unlike in Eq.~\ref{eq:LAPWpotential} planewaves are allowed to enter the muffin-tin spheres in Eq.~\ref{eq:pseudodensity}. The second term is a smooth function that is non-zero only in the muffin-tin spheres. 
$\tilde{\rho}^\alpha(\mathbf{r})	$ are chosen such that $\bar{\rho}(\mathbf{r})$ yields the correct screened potential in the interstitial region.
To ensure it, we expand the second term as follows:
\begin{equation}
\label{eq:pseudomt}
\tilde{\rho}^{\alpha}({\bf r}) = \sum\limits_{l=0}^\infty\sum\limits_{m=-l}^l Q_{lm}^\alpha \ Y_{lm}({\bf \widehat{r}}) \ \sigma_{lm}^{\alpha}(r),
\end{equation}
where $Q_{lm}^\alpha$ are constants chosen in such a way that $\bar{\rho}(\mathbf{r})$ has the same charge multipoles as $\rho(\mathbf{r})$. Since we consider the screened Coulomb interaction, the multipoles {inside the muffin-tin spheres have to be} calculated as 
\begin{equation}
\label{eq:qlm}
q_{lm}^{\alpha} = \frac{(2l+1)!!}{\lambda^{l}}\int_{S_\alpha} Y_{lm}^{*}({\bf
\widehat{r}}) \ i_{l}(\lambda r) \ \rho({\bf r}) d{\bf r},
\end{equation}
where $i_{l}(r)$ is the modified spherical Bessel function of the first kind. 
Finally, the radial functions in Eq.~\ref{eq:pseudomt} are defined as 
\begin{equation}
\sigma_{lm}^{\alpha}(r) = r^{l}\left(1-\frac{r^2}{R_{\alpha}^2}\right)^{n}
\end{equation}
with $n=R_{\alpha}G_{max}/4$. Due to this choice of $\sigma_{lm}^{\alpha}(r)$, the Fourier transform of $\tilde{\rho}^{\alpha}({\bf r})$ can be performed analytically, yielding
\begin{equation}
\fl {\tilde{\rho}^{\alpha}({\bf G}) = \frac{4\pi}{\Omega} e^{-i{\bf G}\cdot{\bf R}^{\alpha}} \sum\limits_{l=0}^\infty\sum\limits_{m=-l}^l \frac{(-i)^{l}}{(2l+1)!!} 
\frac{\lambda^{l+n+1} j_{l+n+1} (G R_{\alpha}) }{i_{l+n+1} (\lambda R_{\alpha}){G}^{n+1}}Y_{lm}({\bf \widehat{G}}) \tilde{q}_{lm}^{\alpha}}.
\end{equation}
With the Fourier transforms of both parts in Eq.~\ref{eq:pseudodensity}, the potential in the interstitial reads 
\begin{equation}
V_{I}({\bf G})=  \frac{4\pi}{G^2+\lambda^2}\,\,\bar{\rho}({\bf G}).
\end{equation}

To obtain the potential in the muffin-tin region, we solve the Dirichlet boundary-value problem with the original density $\rho(\mathbf{r})$. Employing the Green-function method~\cite{Jackson}, the potential reads
\begin{equation}
\label{eq:mtsolution}
V^{\alpha}({\bf r}) = \int_{MT_\alpha} G({\bf r}, {\bf r'})  \rho({\bf r'}) d{\bf r'} - \frac{R_{\alpha}^2}{4\pi} \oint_{S_\alpha}V_{I}(R_{\alpha})  \frac{\partial G}{\partial n'} d\Omega',
\end{equation}
where the Green function and its normal derivative are expressed as
\begin{equation}
\fl G({\bf r}, {\bf r'}) = 4\pi\lambda\sum\limits_{l=0}^\infty\sum\limits_{m=-l}^l i_{l}(\lambda r_{<})k_{l}(\lambda r_{>}) \left[1-\frac{i_{l}(\lambda r_{>})k_{l}(\lambda R_{\alpha})}{k_{l}(\lambda r_{>})i_{l}(\lambda R_{\alpha})}\right]Y_{lm}^{*}({\bf \widehat{r}'})Y_{lm}({\bf \widehat{r}})
\end{equation}
and
\begin{equation}
\label{derivativeG}
\eqalign{\frac{\partial G}{\partial n'} & =\frac{\partial G}{\partial r'} \bigg|_{r'=R_{\alpha}} \\
&=-\frac{4\pi}{R_{\alpha}^2} \sum\limits_{l=0}^\infty\sum\limits_{m=-l}^l Y_{lm}^{*}({\bf \widehat{r}'})Y_{lm}({\bf \widehat{r}}) \frac{i_{l}(\lambda r)}{i_{l}(\lambda R_{\alpha})}},
\end{equation}
respectively. $k_{l}(r)$ is the modified spherical Bessel function of the second kind, and $r_{<}$ ($r_{>}$) is the maximum (minimum) value between ${\bf r}$ and ${\bf r'}$. $\frac{\partial G}{\partial n'}$ is the normal derivative at the atomic sphere boundary.

The method described here is not limited to mixing potentials. The input and output potentials in Eq.~\ref{eq:kerkerRS} can be replaced by respective densities, which does not change the procedure of solving the screened Poisson equation.

\section{Pulay mixing\label{sec:pulay}}

The Pulay mixer~\cite{Pulay1980, Pulay1982} also known as {\it direct inversion in the iterative subspace} represents an improvement over the linear mixer. It uses a sequence of input potentials $v^\mathrm{in}_i(\mathbf{r})$ and residuals $R_i(\mathbf{r})=v^\mathrm{out}_i(\mathbf{r})-v^\mathrm{in}_i(\mathbf{r})$ of previous iterations to estimate the optimal potential and the corresponding residual in the following manner:
\begin{equation}
\label{optimal}
\eqalign{v_{\mathrm{opt}}^{\mathrm{in}}=\Sigma_i\omega_i v_i^{\mathrm{in}}\\
R_{\mathrm{opt}}^{\mathrm{in}}=\Sigma_i\omega_i R_i^{\mathrm{in}}}
\end{equation}
with the weights $\omega_i$ subject to the constraint
\begin{equation}
\label{constraint}
\Sigma_i\omega_i=1.
\end{equation}
Minimizing the residual $R_{\mathrm{opt}}^{\mathrm{in}}$, the weights $\omega_i$ are determined by \cite{Kresse19961, Kresse19962}
\begin{equation}
\label{Pulayweight}
\omega_i=\frac{\Sigma_j A_{ji}^{-1}} {\Sigma_{jk} A_{kj}^{-1}}
\end{equation}
with
\begin{equation}
\label{eq:A}
{A_{ij}=\int R_j^{\mathrm{in}}(\mathbf{r}) R_i^{\mathrm{in}}(\mathbf{r}) d^3r}.    
\end{equation}
According to the conventional Pulay mixing scheme, the next trial potential is given as
\begin{equation}
\label{eq:Pulay}
v_{i+1}^{\mathrm{in}}= v_{\mathrm{opt}}^{\mathrm{in}}+\alpha R_{\mathrm{opt}}^{\mathrm{in}},
\end{equation}
where $\alpha$ is the mixing parameter. The optimal value of $\alpha$ depends on the system \cite{Pulay1982}. For example, it is typically bigger in the case of semiconductors and insulators than in the case of metallic systems.

Equation~\ref{eq:A} contains an integral over the entire unit cell consisting of contributions from the interstitial region and the muffin-tin spheres. When $v_i^{\mathrm{in}}(\mathbf{r})$ in the interstitial region is noisy (e.g. in case the GGA is employed), this noise propagates into the calculated weights resulting in a numerically problematic potential in Eq.~\ref{optimal}. We solve this issue by calculating Eq.~\ref{eq:A} as
\begin{equation}
\label{eq:Amt}
{A_{ij}=\sum\limits_\alpha \int\limits_\mathrm{MT_\alpha} R_j^{\mathrm{in}}(\mathbf{r}) R_i^{\mathrm{in}}(\mathbf{r}) d^3r}.    
\end{equation}
In other words, we ignore the interstitial part of the residuals at this stage and, thus, $R_{\mathrm{opt}}^{\mathrm{in}}$ is minimized strictly within the atomic spheres only. The summation in Eq.~\ref{optimal} is, however, still performed over entire unit cell. The level of noise in the charge density is typically noticeably lower compared to that in the potential, and, hence, employing Eq.~\ref{eq:Amt} is not useful in case of density mixing. Therefore, we use different equations, i.e. Eqs.~\ref{eq:A} and \ref{eq:Amt}, for density and potential mixing, respectively. In both cases, we refer to the method as {\it simple Pulay mixing}.

Previous studies~\cite{Kresse19961, Kresse19962} have shown that introducing the inverse Kerker metric in Eq.~\ref{eq:A} helps to prevent charge sloshing. The matrix elements in this approach read
{
\begin{equation}
\label{eq:Kerkermetric}
A_{ij}=\sum\limits_\mathbf{G} R^{\ast,\mathrm{in}}_i(\mathbf{G}) \frac{G^2+\lambda'^2}{G^2} R_j^{\mathrm{in}}(\mathbf{G}).
\end{equation}
}
Similarly to Eq.~\ref{eq:kerker}, this equation is written in the planewave basis and is thus not directly applicable in FP-LAPW calculations. We solve this issue analogously to the case of the Kerker mixer: 
$R_j(\mathbf{G})/G^2$ corresponds to the bare Coulomb potential due to the charge density $\rho(\mathbf{r})=-R_j(\mathbf{G})/4\pi$.
Finding such a potential is a standard problem in an FP-LAPW code, which is routinely solved using Weinert's method~\cite{Weinert1981}. The expression for the matrix elements can be formally written in real space:
{
 \begin{equation}
 \label{eq:metricreal}
 A_{ij}=\int\int R_i^{\mathrm{in}}(\mathbf{r}) \left[\delta(\mathbf{r}-\mathbf{r^\prime})-\frac{\lambda'^2}{4\pi}\frac{1}{|\mathbf{r}-\mathbf{r^\prime}|}\right] \ R_j^{\mathrm{in}}(\mathbf{r^\prime})  d^3r^\prime d^3r.    
 \end{equation}
 }
The inner integral is evaluated over the entire space. If the range of the outer integral is the whole unit cell, Eq.~\ref{eq:metricreal} is equivalent to Eq.~\ref{eq:Kerkermetric}. We use it for mixing densities. 

The Kerker transformation can also be used as a preconditioner in the Pulay method \cite{Kresse19961, Kresse19962, Shiihara2008, Sundararaman2017}. In this case, the Kohn-Sham potential is updated as
\begin{equation}
\label{eq:Pulay-KP}
v_{i+1}^{\mathrm{in}}(\mathbf{G})= v_{\mathrm{opt}}^{\mathrm{in}}(\mathbf{G})+\alpha\frac{G^2}{G^2+\lambda^2} R_{\mathrm{opt}}^{\mathrm{in}}(\mathbf{G}).
\end{equation}
The expression on the right-hand side exactly matches Eq.~\ref{eq:kerker}, and it can be evaluated using the same procedure as described in Sec.~\ref{sec:kerker}. We use this preconditioner only in the first $n$ iterations (typically, $n=5$). Afterwards, Eq.~\ref{eq:Pulay} is employed.
This modification of the Pulay method with the inverse Kerker metric and the Kerker preconditionning (termed {\it Pulay-KP}) is the most successful method as we will show in the example below.

Both methods, the simple Pulay method and the one with the Kerker preconditioner and the inverse Kerker metric, depend on the mixing parameter $\alpha$ that we set to 0.4 for all considered cases. The latter method contains also the screening parameter $\lambda$, which strongly influences the convergence properties of the method. Previous studies~\cite{Bendt1982, Zhou2018} have suggested to use the wavevector of the Thomas-Fermi screening, $k_{TF}$. Following Ref.~\cite{Zhou2018}, we estimate it as
\begin{equation}
k_{TF}^2 \approx 4\pi N(\varepsilon_F),
\end{equation}
where $N(\varepsilon_F)$ corresponds to the density of states at the Fermi energy.

\begin{figure}
\begin{center}
\includegraphics[width=0.7 \linewidth]{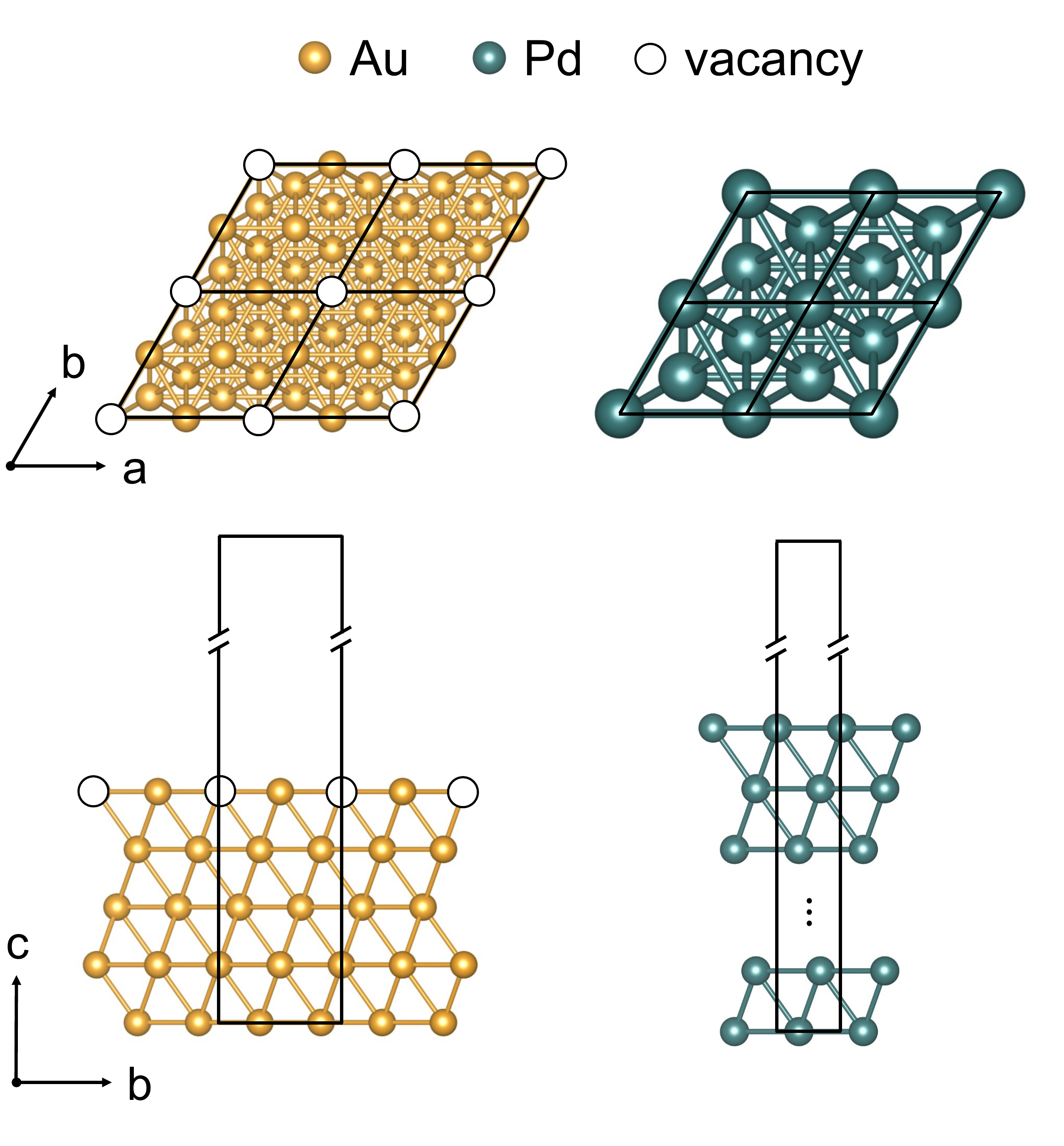}
\end{center}
\caption{Left: Top view (top) and side view (bottom) of a $2\times 2$ Au(111) surface with a vacancy. Right: same for a Pd(111) surface with 15 layers. The black lines indicates the unit cells.} 
\label{fig:AuPdstructure}
\end{figure} 

\section{Computational details}
The first system studied here is the $2 \times 2$ Au(111) surface with a vacancy. 
It consists of 5 layers of Au and is termed 5L-(2 $\times$ 2)Au(111)-V. The second example is the Pd(111) surface with 15 atomic layers, termed 15L-Pd(111). Both structures are displayed in Fig.~\ref{fig:AuPdstructure}. The corresponding slabs are constructed based on the bulk geometries with the lattice constants of 4.19~\AA\ and 3.95~\AA\ for Au and Pd, respectively. These structural parameters are obtained using the PBE exchange-correlation functional. Further structural optimizations of the slabs are not performed. In order to eliminate spurious interactions between the periodic images of the metal slabs, the vacuum spacing in the \textit{z} direction is set to 30~\AA\ and 20~\AA\ for both surfaces.

The above described methods have been implemented in the all-electron full-potential code \exciting\ \cite{Gulans2014}, which is then used for all calculations. It employs the linearized augmented planewave plus local orbitals basis-set (LAPW+lo) \cite{Singh, Sjostedt2000, Madsen2001}. The chosen muffin-tin radii are $R_{MT}^{Au}=$~2.4 bohr and $R_{MT}^{Pd}=$~1.9 bohr. An LAPW cutoff of $R_{MT}G_{max}=$~7 is adopted in all considered systems. Exchange and correlation effects are described within the GGA using the Perdew-Berke-Ernzerhof (PBE) parametrization \cite{Perdew1996}. The Perdew-Wang parametrization~\cite{Perdew1992} of the local-density approximation (LDA) is applied for the sake of comparison. {The Kohn-Sham potential is evaluated with a planewave cutoff of $G^{v}_{max}=$~18~bohr$^{-1}$.} The Brillioun zone (BZ) is sampled on a 4~$\times$~4~$\times$~1 and a 10~$\times$~10~$\times$~1 \textbf{k}-mesh for 5L-(2 $\times$ 2)Au(111)-V and 15L-Pd(111), respectively. The self-consistency cycle is considered converged once the total-energy difference between two consecutive steps is smaller than $10^{-6}$~Ha. Simple Pulay, Pulay with Kerker preconditioner and inverse Kerker metric, and {a variant of the multisecant Broyden method, termed \textit{msec},} employ stored potentials (densities) and residuals obtained from $m=~12$ previous iterations for the 5L-(2 $\times$ 2)Au(111)-V system. In 15L-Pd(111) calculations, we have set $m~=30$, since it is required for a stable performance.

\begin{figure}
\begin{center}
\includegraphics[width=0.7 \linewidth]{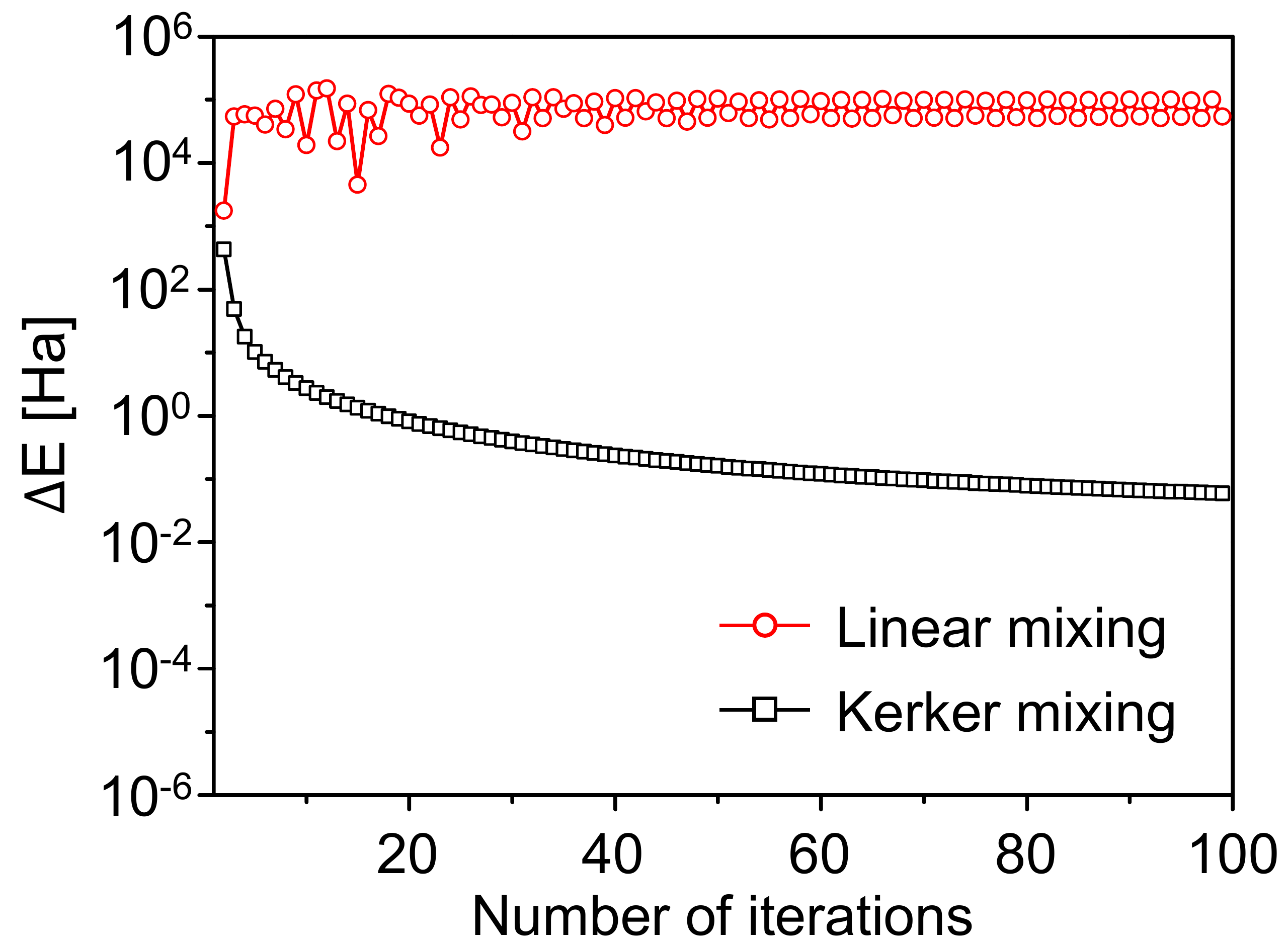}
\end{center}
\caption{Convergence of total energy (in Ha) for 5L-(2 $\times$ 2)Au(111)-V using linear and Kerker mixing schemes.} 
\label{fig:linkerker}
\end{figure}

\section{Results and discussion}
We start discussing the performance of the implemented mixing methods with the case of 5L-(2 $\times$ 2)Au(111)-V.
This is a metallic system with a large unit cell and, therefore, it is prone to the charge-sloshing instability~\cite{Anglade2008, Kerker1981, Kresse19961, Kresse19962, Kohyama1996, Shiihara2008, Sundararaman2017}. Indeed, 5L-(2 $\times$ 2)Au(111)-V is a too complicated problem for the linear mixing and the Kerker method. To illustrate this, we perform calculations applying these methods by mixing the potentials and set the corresponding parameters in Eqs.~\ref{eq:linear} and \ref{eq:kerker} to $\alpha=$~0.4 and $\lambda=1.0$~bohr$^{-1}$. To monitor the convergence, we display in Fig.~\ref{fig:linkerker} the variation of the total energy between two consecutive steps, $\Delta E$. With the linear mixer, this quantity not only does not converge during the first 100 iterations to a desired threshold, but remains large overall, i.e. within range of 10$^{4}$--10$^{5}$ Ha. 
{The huge values of $\Delta E$ reflect excessive changes in the Kohn-Sham potential.
$v_\mathrm{KS}(\mathbf{r})$ varies so much in the muffin-tin region that it heavily affects the radial functions in the LAPW basis.
They acquire a different nodal structure and, hence, trigger enormous fluctuations of the total energy.}
The Kerker mixing, in turn, shows an improved performance compared to the linear mixing. $\Delta E$ decays with the number of iterations, yet this happens too slowly for the Kerker mixer to be practical. Still, we acknowledge that it suppresses charge sloshing. 

\begin{figure}
\begin{center}
\includegraphics[width=0.7 \linewidth]{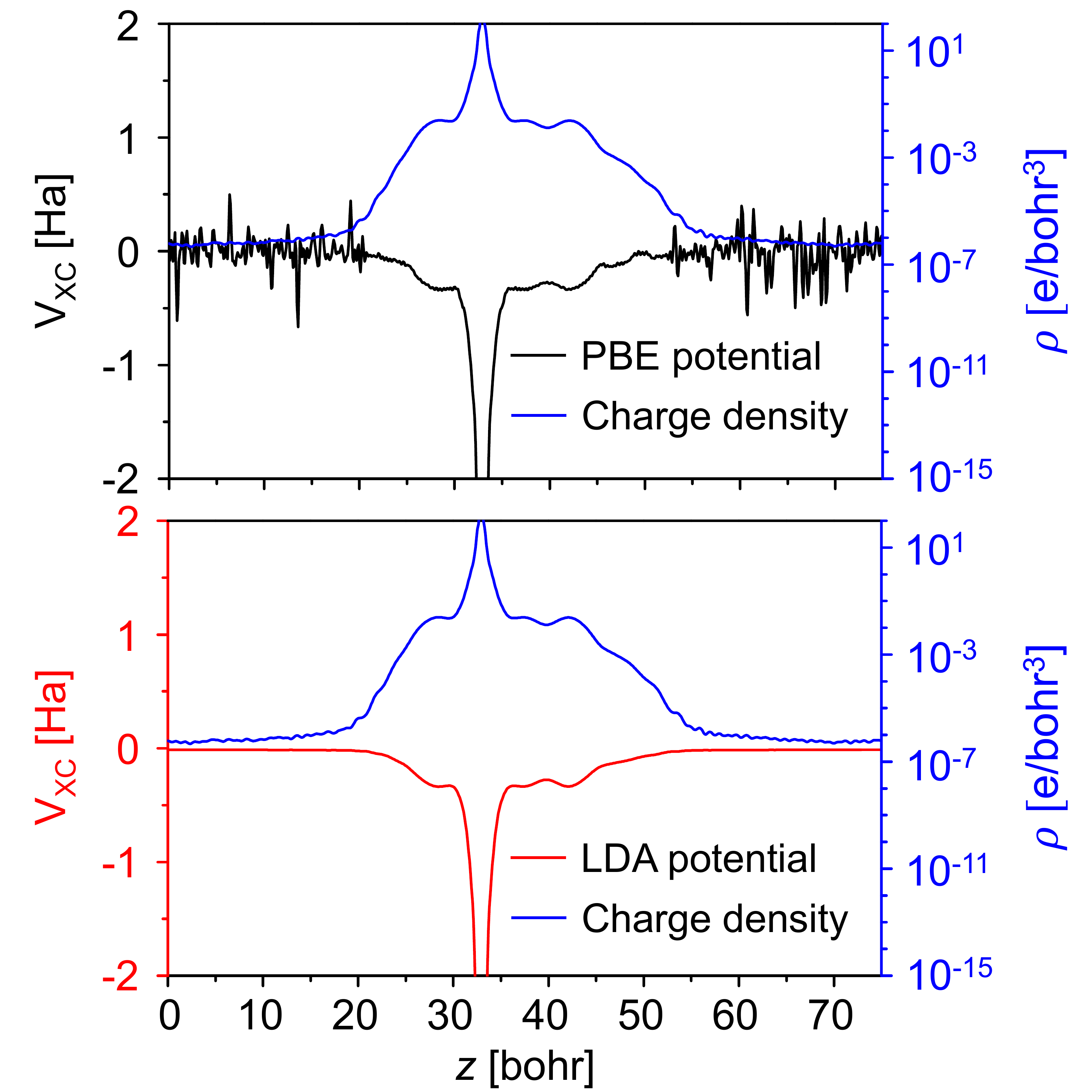}
\end{center}
\caption{Exchange-correlation potential and charge density (blue lines) of 5L-(2 $\times$ 2)Au(111)-V perpendicular to the surface. The exchange-correlation potential from PBE (top) and LDA (bottom) are shown as black and red lines, respectively.} 
\label{fig:vxc}
\end{figure}

A further issue that makes 5L-(2 $\times$ 2)Au(111)-V challenging is the (typically) noisy GGA potential in the vacuum region. To demonstrate this fact, we compare in Fig.~\ref{fig:vxc} the self-consistent exchange-correlation potentials $v_\mathrm{xc}(\mathbf{r})$ obtained by LDA and GGA, together with the corresponding electron densities, for 5L-(2 $\times$ 2)Au(111)-V. While the densities are essentially indistinguishable on the displayed scale, the differences in the exchange-correlation potentials are immediately apparent. The LDA potential is smooth throughout the unit cell and only slighty jagged in the low-density region ($\rho<10^{-4}$~bohr). The noise obviously stems from the gradient term ($\nabla \rho/\rho^{4/3}$), which is prone to rapid variations in low-density regions \cite{Londero2011} since a small variation in $\rho(\mathbf{r})$ can cause a large variation in $v_\mathrm{xc}(\mathbf{r})$. 
{Our numerical tests show that this issue persists even if we increase the LAPW cutoff to $R_{MT}G_{max}=$~8.5 and the planewave cutoff for the potential to $G^{v}_{max}=$~32~bohr$^{-1}$. 
Thus, GGA calculations of low-dimensional systems, i.e. with vacuum, are less stable than those using the  LDA, in particular when mixing potentials.}

\begin{figure}
\begin{center}
\includegraphics[width=0.65 \linewidth]{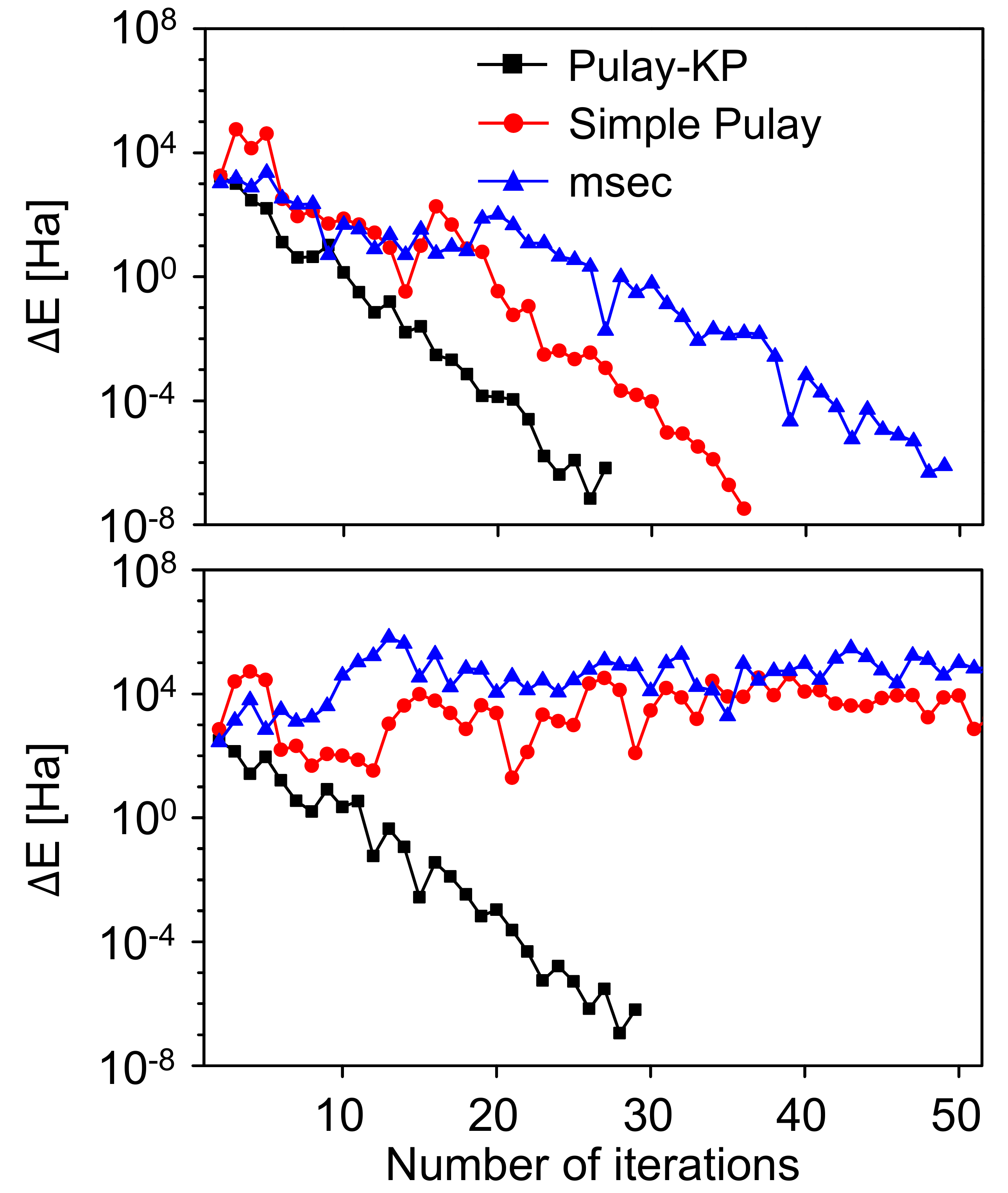}
\end{center}
\caption{Convergence of total energy (in Ha) for 5L-(2 $\times$ 2)Au(111)-V with {msec}, simple Pulay, and Pulay with Kerker preconditioner and inverse Kerker metric (Pulay-KP) methods by using potential (top) and density (bottom) mixing.}
\label{fig:AuPulayKPlinMB}
\end{figure}

Our implementation of both the simple Pulay and the modified Pulay approaches is insensitive to both of these instabilities. This is demonstrated in Fig.~\ref{fig:AuPulayKPlinMB} (top panel) where the convergence behavior of the total energy (using the potential mixing) is compared for these two methods and 
{msec scheme. We note that msec does not include ``unpredicted components" for controlling step size and ``scaling" of planewaves as well as radial functions, compared to a method developed by Marks and Luke \cite{Marks2008}. If these elements were considered, the convergence is expected to be more faster and stable.}
All three methods {employed here}, converge to the target precision within 25--50 steps, in contrast to the linear mixing and the Kerker method (see Fig. \ref{fig:linkerker}). Simple Pulay and Pulay-KP perform better than {the msec method}, where the noise in the potential is not taken care of. 

We also apply these approaches for mixing densities. The convergence behavior of the total energy is shown in the bottom panel of Fig.~\ref{fig:AuPulayKPlinMB}. The simple Pulay and {msec} methods do not reach self-consistency within 100 steps, while the Pulay-KP method converges in just a few iterations more than in the case of potential mixing. Overall, our calculations for 5L-(2 $\times$ 2)Au(111)-V show that the Pulay-KP method is the most efficient method among the considered ones.

\begin{figure}
\begin{center} 
\includegraphics[width=0.65 \linewidth]{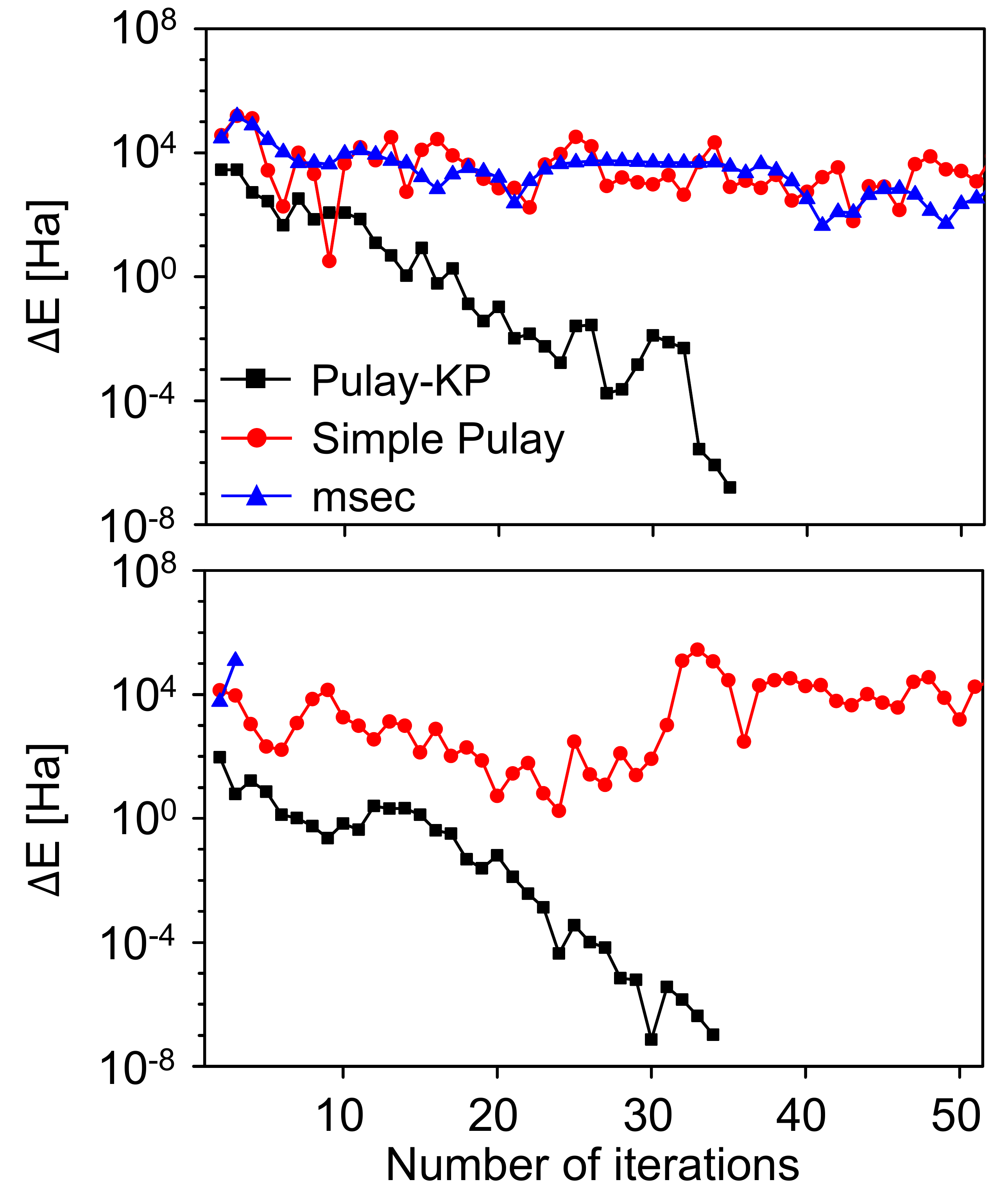}
\end{center}
\caption{Convergence of total energy (in Ha) for 15L-Pd(111) with {msec}, simple Pulay, and Pulay-KP methods by using potential (top) and density (bottom) mixings.}
\label{fig:PdPulayKPlinMB}
\end{figure}

Our second benchmark case is 15-layer Pd(111). Besides being metallic and having a large unit cell, this system has a high density of states near the Fermi level, therefore charge-sloshing is even more pronounced \cite{Woods2019, Mohr2018}. The total-energy convergence in calculations using different approaches, i.e. {msec}, simple Pulay, and Pulay-KP is presented in Fig. ~\ref{fig:PdPulayKPlinMB}. Regardless whether the density or the potential is mixed, the only method that succeeds for this system is Pulay-KP. The other two methods do not converge for either kind of mixing. In fact, {msec} leads in case of density mixing to such an unphysical density that the self-consistency loop cannot be terminated. Pulay-KP, however, reaches the target precision for the total energy in 35 (density) and 36 (potential) steps, respectively. This further indicates that Pulay-PK successfully attenuates the long-wavelength instability which induces charge sloshing.

\section{Summary and Conclusions}
In summary, we have reformulated the Kerker preconditioner and the inverse Kerker metric to make them applicable in FP-LAPW calculations. We have implemented the Pulay mixing algorithm modified with these features in the electronic-structure code \exciting. Our applications demonstrate that this method is robust and superior to the standard Pulay method and the {msec} approach.

\section*{Acknowledgements}
Work supported by the European Community's Horizon 2020 research and innovation program under Marie Sk{\l}odowska-Curie grant agreement No. 675867. Partial support from the the Deutsche Forschungsgemeinschaft (DFG), Projektnummer 182087777 - SFB 951 is acknowledged. All data are stored in the NOMAD Repository, http://dx.doi.org/10.17172/NOMAD/2020.03.25-1. Details on the calculations can be looked up there.

\end{document}